\documentclass[%
 reprint,
 amsmath,amssymb,
 aps,
 pre
]{revtex4-1}
\usepackage{CJK}

\usepackage{amsthm}

\usepackage{bm}

\newtheorem{definition}{Definition}
\newtheorem{thm}{Theorem}
\newtheorem{lem}{Lemma}
\newtheorem{ass}{Assumption}

\theoremstyle{definition}
\newtheorem*{proof*}{Proof}

\def \pw {\left(\omega\right)}

\begin{document}
\begin{CJK*}{UTF8}{gbsn}


\title{The Mathematics of the Ensemble Theory}

\author{Xiang Gao (高翔)}
 \email{qasdfgtyuiop@gmail.com}
\affiliation{%
6889 Rochelle Ave, Newark, CA 94560, United States
}%

\date{\today}

\begin{abstract}
This study shows that the generalized Boltzmann distribution is the only distribution mathematically consistent with thermodynamics when the system is described by an ensemble of a certain mathematical form. This mathematical form is very general, such that the canonical, grand-canonical, or isothermal-isobaric ensemble theories are all special cases of this form. Compared with the standard textbook formalism of the statistical mechanics (SM), this approach does not require a prior distribution, does not assume the functional form or maximization of entropy, and employs fewer assumptions. Therefore, this new insight challenges the belief on the requirement of a prior distribution in SM and provides a new way to derive the Boltzmann distribution. This study also reveals the logical and mathematical constraints of SM's fundamental components; therefore, it could potentially benefit researchers on non-Boltzmann-Gibbs SM and philosophers studying the foundations of SM.
\end{abstract}


\maketitle

\end{CJK*}


\section{Introduction}

The statistical mechanics(SM) dates back to Boltzmann\cite{Boltzmann2012} and Gibbs\cite{JosiahWillardGibbs1902} and has become the core of modern physics for describing matters and radiations. The role that SM plays in modern physics is to fill the gap between thermodynamics and microscopic theories (classical or quantum mechanics). Although it is widely believed that given the current microscopic state of a system, it is possible to predict the microscopic state of that system at any time in the future, there is no known method to derive thermodynamics from only microscopic theories. Indeed, even defining concepts in thermodynamics is already very tricky:

Consider an isolated system $\mathcal{S}$ with $N$ particles whose dynamics are given by classical mechanics. Alice has some superpower to know the positions $q_1, \ldots, q_{3N}$ and momentums $p_1, \ldots,p_{3N}$ of all the $N$ particles in $\mathcal{S}$; that is, Alice knows at which exact point of the $\Gamma$ space the system currently is. How should Alice compute the volume, pressure, temperature, internal energy, entropy, etc., as a function $\Gamma\to\mathbb{R}$ of $q_1, \ldots, q_{3N}, p_1, \ldots,p_{3N}$? How should Alice judge whether this system is in equilibrium or not, as a function $\Gamma\to\left\{\text{true}, \text{false}\right\}$ of $q_1, \ldots, q_{3N}, p_1, \ldots,p_{3N}$? How can Alice derive the second law of thermodynamics $S\left(q_1, \ldots, q_{3N}, p_1, \ldots,p_{3N}\right) \leq S\left(q'_1, \ldots, q'_{3N}, p'_1, \ldots,p'_{3N}\right)$ for $t'>t$, from classical mechanics?

The ensemble theory was invented to overcome these difficulties. Instead of studying the system at a single microscopic state, the ensemble theory considers the system's microscopic state as unknown and employs a probability density function to describe each state's probability. It is widely believed that the probability density can not be derived from the microscopic theory and has to be obtained from additional assumptions (therefore sometimes they are referred to as ``a prior distribution''). For example, isolated systems are assumed to have uniform distributions; closed systems are assumed to be a small part of a much larger isolated system with uniform distributions. With these assumptions, microcanonical, canonical, grand-canonical, and isothermal-isobaric ensembles are proved to have uniform and generalized-Boltzmann distributions\footnote{The term ``generalized-Boltzmann distributions'' is introduced in \cite{Gao2019}. It refers to the distribution of the form $Pr\left(\omega\right)\propto\exp\left[\sum_{\eta=1}^{n}\frac{X_{\eta}x_{\eta}^{\left(\omega\right)}}{k_{B}T}-\frac{E^{\left(\omega\right)}}{k_{B}T}\right]$} respectively. Further assumptions are required to obtain thermodynamics state functions. For example, the entropy of isolated systems is assumed to be $S=k_B\log\left|\Omega\right|$; closed systems' internal energy is assumed to be the ensemble average $\sum_\omega\Pr(\omega) E^{(\omega)}$.

This article shows that, if a system is described by an ensemble theory of a specific mathematical form, then the generalized-Boltzmann distribution is the only distribution that can reproduce the thermodynamics of that system.  This mathematical form is very general and the canonical, grand-canonical, or isothermal-isobaric ensemble theories are special cases of this mathematical theory. Compared to how textbooks\cite{kardar2007statistical, callen1998thermodynamics, chandler1988introduction, landau2013course, balescu1975equilibrium, tolman1979principles} formalize SM, this approach does not require the assumption of a prior distribution. The set of assumptions in this approach can be considered a strict subset of the assumptions of textbook approaches. With this insight, this article has the following contribution:

First, it challenges the belief that a prior distribution is required in SM. Instead, this article indicates that it is the mathematical form and the consistency with thermodynamics that determine an ensemble's probability density.

Second, it introduces a new method to derive the generalized-Boltzmann distribution. This derivation is only based on the mathematical form of the ensemble. Compared to standard textbook approaches and newer approaches like\cite{Gao2019}, this way to derive the generalized-Boltzmann distribution requires fewer assumptions.

Third, this article could provide new insights into non-Boltzmann-Gibbs (non-BG) SM such as the Tsallis statistics\cite{Tsallis1988, Curado1991, Tsallis1998, Tsallis2019, tsallis2009introduction,boon2005nonextensive}. Non-BG SM is widely used in the study of complex systems. A notable difficulty that non-BG SM researchers are facing is how to construct a self-consistent theory, while at the same time, being consistent with thermodynamics.

Fourth, the conclusion of this article might be helpful for philosophers studying the foundations of SM. Interested readers are referred to \cite{RomanFrigg2011} for a comprehensive field review. Earlier reviews\cite{Haar1955, Penrose1979} might also be helpful.

\section{Definitions, notations, and assumptions\label{sec:ass}}

Before moving to our theory's formal statement, let us first revisit an ensemble theory's essential components:

The first and foremost is the \emph{set of microstates}, denoted by $\Omega$. The set of microstates is the underlying set of the measure space and is determined by the ensemble theory and the underlying microscopic theory together. For example, for a microcanonical ensemble, the set of microstates contains all the points in the $\Gamma$ space of classical mechanics whose coordinates satisfy volume constraints and energy is a constant $E$ or in a small range of energies $\left[E,E+\Delta E\right]$, or all eigenstates of the Hamiltonian operator of quantum mechanics under this energy constraints.

The second is the different roles that different thermodynamic state functions play in the ensemble theory. They are classified as: parameters determining $\Omega$, parameters determining the probability density, quantities associated with random variables, and other statistical quantities. Let us take a look at the canonical ensemble as an example. The volume and the number of particles are parameters determining $\Omega$. These parameters together with the underlying microscopic theory defines which microstates are contained in $\Omega$, but they do not directly appear in the equation of the probability density. The temperature is also a parameter; this parameter has nothing to do with $\Omega$, but it appears as a parameter of the probability density function. The system's energy is a quantity associated with random variable: for each microstate, there is corresponding energy. Other statistical quantities include pressure and entropy. Since neither $\Omega$ nor the probability density directly depend on these two quantities, they are not parameters. They are not quantities associated with random variables either because a single microstate does not have a well-defined pressure or entropy.

The third is the probability density. It is a function of random variables and parameters determining the probability density, but not the other quantities. For the example of the canonical ensemble, the probability density is a function $f\left(E,T\right)$ of $E$ and $T$, but not $N$, $V$, $S$, $p$ directly.

The last but not the least is the set of rules connecting each random variable with its corresponding thermodynamic state functions. For the case of the canonical, grand-canonical, or isothermal-isobaric ensembles, these rules are $U=\sum_\omega\Pr(\omega)E^{(\omega)}$, $N=\sum_\omega\Pr(\omega)N^{(\omega)}$, etc., but for the case of the Tsallis statistics, the rule is more complicated\cite{boon2005nonextensive}.

Being aware of these essential components, we are now ready to define the mathematical form of our theory formally:

\begin{definition}\label{def:system}
In this article, we will study a thermodynamic system with generalized forces $X_{1},\ldots,X_{n}$,
$Y_{1},\ldots,Y_{m}$ and generalized coordinates $\chi_{1},\ldots,\chi_{n}$,
$y_{1},\ldots,y_{m}$. The first law of thermodynamics for this system
states that
\begin{equation}
dU=TdS+\sum_{\eta=1}^{n}X_{\eta}d\chi_{\eta}+\sum_{\eta=1}^{m}Y_{\eta}dy_{\eta}\label{eq:1st-law}.
\end{equation}
We want to describe this thermodynamic system with an ensemble parametrized by $T,X_{1},\ldots,X_{n},y_{1},\ldots,y_{m}$, where $y_{1},\ldots,y_{m}$ determines the set of microstates, and $T,X_{1},\ldots,X_{n}$ determines the probability density.
In this setup, $E,x_{1},\ldots,x_{n}$ are random variables; their corresponding thermodynamic state functions are $U,\chi_{1},\ldots,\chi_{n}$. $Y_{1},\ldots,Y_{m}, S$ are other statistical quantities of that ensemble. For a microstate $\omega\in\Omega$,
we denote the value of random variables at $\omega$ by $E^{\pw},x_{1}^{\pw},\ldots,x_{n}^{\pw}$.
For clarity, we have used $X$ vs $Y$ to distinguish ensemble parameters
from statistical quantities. In addition, we utilize $\chi$ vs $x$ and $U$ vs $E$ to distinguish thermodynamic state functions from random variables. Specific heats of the system are assumed to be positive\footnote{There do exist systems with negative specific heat. These systems have odd properties. For example, they are never extensive, and they cannot achieve thermal equilibrium with a large heat bath. Discussion of such systems is beyond the scope of this article. Interested readers are referred to \cite{lynden1999negative}}.
\end{definition}

The mathematical form stated in definition \ref{def:system} is very general, and the canonical, grand-canonical, or isothermal-isobaric ensemble theories are all special cases of this form. Let us take a look at a few examples:
\begin{enumerate}
\item The canonical ensemble of a single component system is also called an $NVT$ ensemble. This ensemble has $y_1=N$, $y_2=V$, $Y_1=\mu$, $Y_2=-p$. There are no $\chi$, $X$ or $x$. The first law for this system reads $dU=TdS-pdV+\mu dN$.
\item The grand canonical ensemble of a two component system is also called a $\mu_1\mu_2 V T$ ensemble. This ensemble has $y_1=V$, $Y_1=-p$, $\chi_1=N_1$, $\chi_2=N_2$, $X_1=\mu_1$, $X_2=\mu_2$, $x_1^{\left(\omega\right)}=N_1^{\left(\omega\right)}$, and $x_2^{\left(\omega\right)}=N_2^{\left(\omega\right)}$. The first law for this system reads $dU=TdS-pdV+\mu_1 dN_1+\mu_2 dN_2$.
\item The isothermal-isobaric ensemble of a single component is also called an NpT ensemble. It has $y_1=N$, $Y_1=\mu$, $\chi_1=V$, $X_1=-p$, and $x_1^{\left(\omega\right)}=V^{\left(\omega\right)}$. The first law for this system reads $dU=TdS-pdV+\mu dN$.
\end{enumerate}

In textbooks\cite{kardar2007statistical, callen1998thermodynamics, chandler1988introduction, landau2013course, balescu1975equilibrium, tolman1979principles}, the probability density $\Pr\left(\omega\right)$ (i.e. the generalized Boltzmann distribution) of the ensemble defined in definition \ref{def:system} is usually derived by assuming:
\begin{enumerate}
\item The microcanonical ensemble has a uniform distribution.
\item The entropy of a microcanonical ensemble is given by $S=k_B\log\left|\Omega\right|$
\item The ensemble defined in definition \ref{def:system} can be considered as a system in equilibrium with a reservoir. The interaction between the system and the reservoir is weak. Furthermore, the system, together with the reservoir as a whole, can be described by a microcanonical ensemble. Or, alternatively,
\item The entropy $S=-k_B \sum_\omega \Pr(\omega)\log\Pr(\omega)$ is maximal with respect to the probability density function $ \Pr(\omega)$ under the constraints that $U=\sum_\omega\Pr(\omega)E^{\left(\omega\right)}$,
$\chi_{1}=\sum_\omega\Pr(\omega) x_{1}^{\left(\omega\right)}$, $\cdots$,
$\chi_{n}=\sum_\omega\Pr(\omega) x_{n}^{\left(\omega\right)}$ being constant.
\end{enumerate}

This article employs a different set of assumptions. The reader will soon find that these assumptions are just a subset of standard approaches in textbooks or common sense. We will show that we can derive the generalized Boltzmann distribution using this small subset of standard approaches or common sense.

\begin{ass}\label{ass:propto}
The probability density function $\Pr\pw$ is proportional to a function
\begin{equation}
\Pr\pw\propto f\left(E^{\pw},x_{1}^{\pw},\ldots,x_{n}^{\pw};T,X_{1},\ldots,X_{n}\right)
\end{equation}
The function will be denoted by $f\pw$ or $f_{\omega}$ in short.
\end{ass}

This is a standard assumption in textbooks. In textbooks, the system being studied is assumed to be in contact with a reservoir. The interaction between the system and the reservoir is assumed to be weak. This weak-interacting assumption means the microstates of system+reservoir are the cartesian product of the microstates of the system and the reservoir. This implies that the probability density is proportional to the number of microstates in the reservoir.

\begin{ass}\label{ass:avg}
Random variables are connected to their corresponding state functions through ensemble average:
\begin{equation}
\begin{array}{c}
U=\sum_\omega\Pr(\omega) E^{\left(\omega\right)} \\
\chi_{1}=\sum_\omega\Pr(\omega) x_{1}^{\left(\omega\right)} \\
\vdots\\
\chi_{n}=\sum_\omega\Pr(\omega) x_{n}^{\left(\omega\right)} 
\end{array}
\end{equation}
\end{ass}

The above assumption is also standard in textbooks. It is not used to derive the generalized Boltzmann distribution, but it is required to obtain thermodynamic state functions after obtaining the distribution. Only with this assumption, people can derive the connection between the partition functions and a thermodynamic state function and then derive the rest state functions taking advantage of natural variables. There exist non-BG statistical mechanics where this assumption does not hold\cite{boon2005nonextensive}.

\begin{ass}\label{ass:Tinc}
At infinite temperature, all the microstates have the same probability.
\end{ass}

People usually consider this assumption common sense instead of writing it out in textbooks. In the author's opinion, this assumption should be viewed as a qualitative definition of infinite temperature. 

It is worth mentioning that the only assumption about entropy this article made is that the entropy is a state function satisfying equation \ref{eq:1st-law} in definition \ref{def:system}. The functional form of entropy is not assumed. Instead, the functional form of entropy is a conclusion of theorem \ref{fake-dist-thm} as shown in theorem \ref{entropy} and its proof. This article does not assume the entropy is being maximized either.

This article does not assume the form of the underlying microscopic theory, so the conclusions of this article should fit well in both classical mechanics and quantum mechanics.

\section{The Theory}

The main result of this article is the following theorem.
\begin{thm}
\label{fake-dist-thm}
An ensemble as defined in definition \ref{def:system} and satisfies assumptions \ref{ass:propto}, \ref{ass:avg}, and \ref{ass:Tinc} obeys the generalized Boltzmann distribution:
\begin{equation}
\Pr\pw\propto\exp\left[\sum_{\eta=1}^{n}\frac{X_{\eta}x_{\eta}^{\pw}}{k_{B}T}-\frac{E^{\pw}}{k_{B}T}\right]\label{eq:gen-boltzmann}.
\end{equation}
\end{thm}

\begin{proof*}
Our proof employs some lemmas, that are stated and proved in section \ref{sec:lemmas}. From assumption \ref{ass:propto}, we can write $\Pr\pw$ as follows:
\begin{equation}
\Pr\pw\propto f\left(E^{\pw},x_{1}^{\pw},\ldots,x_{n}^{\pw};T,X_{1},\ldots,X_{n}\right)
\end{equation}
Let $\beta=\frac{1}{k_{B}T}$,
$\tilde{X}_{\eta}=\beta X_{\eta}$ and $\tilde{Y}_{\eta}=\beta Y_{\eta}$.
Instead of writing $f$ as a function of $\left(T,X_{1},\ldots,X_{n}\right)$,
we will write it as a function of $\left(\beta,\tilde{X}_{1},\ldots,\tilde{X}_{n}\right)$:
\begin{equation}\label{eq:propto-general}
\Pr\pw\propto f\left(E^{\pw},x_{1}^{\pw},\ldots,x_{n}^{\pw};\beta,\tilde{X}_{1},\ldots,\tilde{X}_{n}\right)
\end{equation}

Rewrite the first law of thermodynamics (equation \ref{eq:1st-law}) with $\beta,\tilde{X}_{1},\ldots,\tilde{X}_{n}$,
we get
\begin{equation}
\frac{dS}{k_{B}}=\beta dU-\sum_{\eta=1}^{n}\tilde{X}_{\eta}d\chi_{\eta}-\sum_{\eta=1}^{m}\tilde{Y}_{\eta}dy_{\eta}
\end{equation}
do Legendre transformation to get a state function $B$ with natural
variables $\beta$, $\tilde{X}_{1},\ldots,\tilde{X}_{n}$, $y_{1},\ldots,y_{m}$,
we have 
\begin{equation}
B=\frac{S}{k_{B}}-\beta U+\sum_{\eta=1}^{n}\tilde{X}_{\eta}\chi_{\eta}\label{eq:defb}
\end{equation}
\begin{equation}
dB=-Ud\beta+\sum_{\eta=1}^{n}\chi_{\eta}d\tilde{X}_{\eta}-\sum_{\eta=1}^{m}\tilde{Y}_{\eta}dy_{\eta}\label{eq:diffB}
\end{equation}
therefore
\begin{equation}
U=\sum_\omega\Pr(\omega) E^{\pw} =-\frac{\partial B}{\partial\beta}\label{eq:eb}
\end{equation}
\begin{equation}
\chi_{\eta}=\sum_\omega\Pr(\omega)x_{\eta}^{\pw} =\frac{\partial B}{\partial\tilde{X}_{\eta}}\label{eq:xb}
\end{equation}
The normalization constant (partition function) for equation \ref{eq:propto-general} is
\begin{equation}
Z=\sum_{\omega}f_{\omega}\label{eq:partition_function}
\end{equation}
where $f_{\omega}$ is short for
\begin{equation}
f\left(E^{\pw},x_{1}^{\pw},\ldots,x_{n}^{\pw};\beta,\tilde{X}_{1},\ldots,\tilde{X}_{n}\right)
\end{equation}
Then equation \ref{eq:eb} and equation \ref{eq:xb} becomes
\begin{equation}
\sum_{\omega}\frac{E^{\pw}f_{\omega}}{Z}=-\frac{\partial B}{\partial\beta}
\end{equation}
\begin{equation}
\sum_{\omega}\frac{x_{\eta}^{\pw}f_{\omega}}{Z}=\frac{\partial B}{\partial\tilde{X}_{\eta}}
\end{equation}
From basic multivariable calculus, we have $\frac{\partial^{2}B}{\partial\tilde{X}_{\eta}\partial\beta}=\frac{\partial^{2}B}{\partial\beta\partial\tilde{X}_{\eta}}$.
Therefore
\begin{equation}
\frac{\partial}{\partial\tilde{X}_{\eta}}\sum_{\omega}\frac{E^{\pw}f_{\omega}}{Z}+\frac{\partial}{\partial\beta}\sum_{\omega}\frac{x_{\eta}^{\pw}f_{\omega}}{Z}=0
\end{equation}
which simplifies to
\begin{equation}
\sum_{\omega}\left[E^{\pw}\frac{\partial\left(f_{\omega}/Z\right)}{\partial\tilde{X}_{\eta}}+x_{\eta}^{\pw}\frac{\partial\left(f_{\omega}/Z\right)}{\partial\beta}\right]=0
\end{equation}
the above equality should always be true, regardless of the details
of the system and microstates, the only way to guarantee this is to
have
\begin{equation}
E^{\pw}\frac{\partial\left(f_{\omega}/Z\right)}{\partial\tilde{X}_{\eta}}+x_{\eta}^{\pw}\frac{\partial\left(f_{\omega}/Z\right)}{\partial\beta}=0
\end{equation}
for all $\omega$s. Apply the same thing to $\frac{\partial^{2}B}{\partial\tilde{X}_{i}\partial\tilde{X}_{j}}=\frac{\partial^{2}B}{\partial\tilde{X}_{j}\partial\tilde{X}_{i}}$
and from lemma \ref{f(abcd)}, we know that $f$ must have the form
$g\left(\zeta,E^{\pw},x_{1}^{\pw},\ldots,x_{n}^{\pw}\right)$,
where
\begin{equation}
\zeta=\beta E^{\pw}-\sum_{\eta=1}^{n}\tilde{X}_{\eta}x_{\eta}^{\pw}
\end{equation}
Let $G$ be an antiderivative of $g$ with respect to $\zeta$, that
is,
\begin{equation}\label{eq:G'=g}
G'=g\left(\zeta,E^{\pw},x_{1}^{\pw},\ldots,x_{n}^{\pw}\right)
\end{equation}
We use the prime $'$ exclusively for derivative with respect to the
first argument $\zeta$ while keeping other arguments $E^{\pw},x_{1}^{\pw},\ldots,x_{n}^{\pw}$
constant. Let $K=\sum_{\omega}G\left(\zeta;E^{\pw},x_{1}^{\pw},\ldots,x_{n}^{\pw}\right)$,
it is easy to show that
\begin{equation}
\frac{\partial K}{\partial\beta}=\sum_{\omega}E^{\pw}g_{\omega}=Z\cdot \sum_\omega\Pr(\omega)E^{\pw} =-Z\cdot\frac{\partial B}{\partial\beta}\label{eq:KZB-beta}
\end{equation}
\begin{equation}
\frac{\partial K}{\partial\tilde{X}_{\eta}}=-\sum_{\omega}x_{\eta}^{\pw}g_{\omega}=-Z\cdot\sum_\omega\Pr(\omega) x_{\eta}^{\pw} =-Z\cdot\frac{\partial B}{\partial\tilde{X}_{\eta}}\label{eq:KZB-X}
\end{equation}
where $g_{\omega}$ is short for
\begin{equation}
g\left(\beta E^{\pw}-\sum_{\eta=1}^{n}\tilde{X}_{\eta}x_{\eta}^{\pw},E^{\pw},x_{1}^{\pw},\ldots,x_{n}^{\pw}\right)
\end{equation}
Note that $K,Z,B$ all have the same set of natural variables $\beta$,
$\tilde{X}_{1},\ldots,\tilde{X}_{n}$, $y_{1},\ldots,y_{m}$, so equation \ref{eq:KZB-beta}
and equation \ref{eq:KZB-X} can be condensed as 
\begin{equation}
dK=-Z\cdot dB\label{eq:dK=00003D-ZdB}
\end{equation}
Properties of exact differential requires that $K$, $Z$, and $B$
must have a function relationship between each other. 

Besides, $K$ and $Z$ are both functionals with parameters $\beta,\tilde{X}_{1},\ldots,\tilde{X}_{n}$
that map functions of $\omega$ (random variables $E^{\pw},x_{1}^{\pw},\ldots,x_{n}^{\pw}$)
to numbers. If the random variables change by a small amount $\delta E^{\pw},\delta x_{1}^{\pw},\ldots,\delta x_{n}^{\pw}$,
then these functionals change as follows:
\begin{eqnarray}
\delta K&=&\sum_{\omega}\left[\left(\frac{\partial G}{\partial E^{\pw}} + \beta g_{\omega}\right)\delta E^{\pw}\right.\nonumber\\
&+&\left.\sum_{\eta}\left(\frac{\partial G}{\partial x_{\eta}^{\pw}}-\tilde{X}_{\eta}g_{\omega}\right)\delta x_{\eta}^{\pw}\right]
\end{eqnarray}
\begin{eqnarray}
\delta Z & = & \sum_{\omega}\left[\left(\frac{\partial g}{\partial E^{\pw}}+\beta g'_{\omega}\right)\delta E^{\pw}\right.\nonumber\\
& + & \left.\sum_{\eta}\left(\frac{\partial g}{\partial x_{\eta}^{\pw}}-\tilde{X}_{\eta}g'_{\omega}\right)\delta x_{\eta}^{\pw}\right]
\end{eqnarray}
where the $\frac{\partial}{\partial E^{\pw}}$ and
$\frac{\partial}{\partial x_{\eta}^{\pw}}$ are partial
derivatives keeping $\zeta$ constant:
\begin{equation}
\left.\frac{\partial}{\partial E^{\pw}}\right|_{\zeta,x_{1}^{\pw},\ldots,x_{n}^{\pw}}
\end{equation}
\begin{equation}
\left.\frac{\partial}{\partial x_{\eta}^{\pw}}\right|_{\zeta,E^{\pw},x_{1}^{\pw},\ldots x_{\eta-1}^{\pw},x_{\eta+1}^{\pw},\ldots,x_{n}^{\pw}}
\end{equation}
The function relationship between $K$ and $Z$ requires $\delta K=C\left(\beta,\tilde{X}_{1},\ldots,\tilde{X}_{n}\right)\delta Z$
to be true for all $\delta E^{\pw},\delta x_{1}^{\pw},\ldots,\delta x_{n}^{\pw}$,
where $C\left(\beta,\tilde{X}_{1},\ldots,\tilde{X}_{n}\right)$ is
some constant that must not depend on $E^{\pw},x_{1}^{\pw},\ldots,x_{n}^{\pw}$
but could depend on $\beta,\tilde{X}_{1},\ldots,\tilde{X}_{n}$. Then
\begin{equation}
\frac{\partial G}{\partial E^{\pw}}+\beta g_{\omega}=C\left(\beta,\tilde{X}_{1},\ldots,\tilde{X}_{n}\right)\cdot\left[\frac{\partial g}{\partial E^{\pw}}+\beta g'_{\omega}\right]\label{eq:g=00003Dcg'}
\end{equation}
\begin{equation}
\frac{\partial G}{\partial x_{\eta}^{\pw}}-\tilde{X}_{\eta}g_{\omega}=C\left(\beta,\tilde{X}_{1},\ldots,\tilde{X}_{n}\right)\cdot\left[\frac{\partial g}{\partial x_{\eta}^{\pw}}-\tilde{X}_{\eta}g'_{\omega}\right]\label{eq:G-g=00003DCg-g'}
\end{equation}
From lemma \ref{fgc-constant}, $C\left(\beta,\tilde{X}_{1},\ldots,\tilde{X}_{n}\right)$
is a constant that does not depends on $\beta,\tilde{X}_{1},\ldots,\tilde{X}_{n}$.
Denote it as $C_{1}$. Define 
\begin{equation}
\hat{\mathcal{L}}=\frac{\partial}{\partial E^{\pw}}+\beta\frac{\partial}{\partial\zeta}
\end{equation}
then equation \ref{eq:g=00003Dcg'} can be written as $\hat{\mathcal{L}}G=C_{1}\hat{\mathcal{L}}g$.
Since $\hat{\mathcal{L}}$ is a linear operator, we have $\hat{\mathcal{L}}\left(G-C_{1}g\right)=0$.
The kernel of $\hat{\mathcal{L}}$ contains functions of the form
$\varphi\left(\zeta-\beta E^{\pw}\right)$. By performing a
same thing to equation \ref{eq:G-g=00003DCg-g'}, we see that
$G-C_{1}g$ must have the form 
\begin{equation}
G-C_{1}g=\varphi\left(\zeta-\beta E^{\pw}+\sum_{\eta=1}^{n}\tilde{X}_{\eta}x_{\eta}^{\pw}\right).
\end{equation}
Since $\zeta-\beta E^{\pw}+\sum_{\eta=1}^{n}\tilde{X}_{\eta}x_{\eta}^{\pw}\equiv0$,
we then have that $G-C_{1}g=C_{3}$ where $C_{3}$ denotes another constant.
By taking the derivative of both side, we get that $g=C_{1}g'$, which immediately
leads to
\begin{eqnarray}
g\left(\zeta,E^{\pw},x_{1}^{\pw},\ldots,x_{n}^{\pw}\right)&=&\nonumber\\
C_{2}\left(E^{\pw},x_{1}^{\pw},\ldots,x_{n}^{\pw}\right)&\cdot&\exp\left(\zeta/C_{1}\right)
\end{eqnarray}
and
\begin{eqnarray}
G\left(\zeta,E^{\pw},x_{1}^{\pw},\ldots,x_{n}^{\pw}\right)&=&\nonumber\\
C_{1}\cdot C_{2}\left(E^{\pw},x_{1}^{\pw},\ldots,x_{n}^{\pw}\right)&\cdot&\exp\left(\zeta/C_{1}\right),
\end{eqnarray}
where $C_{2}\left(E^{\pw},x_{1}^{\pw},\ldots,x_{n}^{\pw}\right)$
denotes a constant that must not depend on $\beta,\tilde{X}_{1},\ldots,\tilde{X}_{n}$
but could depend on $E^{\pw},x_{1}^{\pw},\ldots,x_{n}^{\pw}$.
In order to have positive specific heat, we must have $C_{1}<0$.
Since $C_{1}$ denotes a constant multiplied
toward the temperature, from lemma \ref{temp-scale}, we can choose
$C_{1}=-1$ without loss of generality. By defining 
\begin{equation}
\Lambda\pw=C_{2}\left(E^{\pw},x_{1}^{\pw},\ldots,x_{n}^{\pw}\right),
\end{equation}
we have
\begin{equation}
\Pr\pw\propto\Lambda\pw\cdot\exp\left[\sum_{\eta=1}^{n}\frac{X_{\eta}x_{\eta}^{\pw}}{k_{B}T}-\frac{E^{\pw}}{k_{B}T}\right].
\end{equation}

As $T\to\infty$, the probability density function $\Pr\pw \to \Lambda\pw$. From assumption \ref{ass:Tinc}, $\Lambda\pw$ must be a constant. This completes the proof.
\end{proof*}

The procedure to obtain all the other thermodynamic state functions is the same as in textbooks.
In the proof of theorem
\ref{fake-dist-thm}, it is easy to see that $K=-Z$. Applying
equation \ref{eq:dK=00003D-ZdB}, we obtain that $B=\log Z+C_{4}$. $B$
being extensive implies that $C_{4}$ must vanish. Then, we have that $B=\log Z$. Let
\begin{equation}
J=k_{B}T\cdot B=k_{B}T\log Z\label{eq:free_energy}
\end{equation}
and substitute into equation \ref{eq:defb}
and equation \ref{eq:diffB}, we get that
\begin{equation}
J=TS-U+\sum_{\eta=1}^{n}X_{\eta}\chi_{\eta}
\end{equation}
\begin{equation}
dJ=SdT+\sum_{\eta=1}^{n}\chi_{\eta}dX_{\eta}-\sum_{\eta=1}^{m}Y_{\eta}dy_{\eta}\label{eq:dJ}
\end{equation}
Then we can obtain all the state functions by taking advantage of natural variables. Let us take a look at entropy as an example:
\begin{thm}
\label{entropy}\label{thm:entropy}
The ensemble as defined in definition \ref{def:system} and satisfies assumptions \ref{ass:propto}, \ref{ass:avg}, and \ref{ass:Tinc} has entropy:
\begin{equation}
S=-k_B\sum_\omega\Pr\pw\log\Pr\pw\label{eq:entropy}.
\end{equation}
\end{thm}
\begin{proof*}
For brevity, in the context of this proof, we will use $\frac{\partial}{\partial T}$ and $\frac{\partial}{\partial \beta}$ to denote
$\left.\frac{\partial }{\partial T}\right|_{X_1,\ldots,X_n,y_1,\ldots,y_m}$ and $\left.\frac{\partial}{\partial \beta}\right|_{X_1,\ldots,X_n,y_1,\ldots,y_m}$.

From equation \ref{eq:dJ}, we have
\begin{equation}
S=\frac{\partial J}{\partial T}=-\frac{1}{k_B T^2}\frac{\partial J}{\partial \beta}\label{eq:seqpjpbata}
\end{equation}
From equation \ref{eq:free_energy}, we have
\begin{equation}
\frac{\partial J}{\partial \beta} = \frac{1}{\beta^2}\left(\frac{\beta}{Z}\frac{\partial Z}{\partial \beta}-\log Z\right).\label{eq:pjpbeta}
\end{equation}
Let $\gamma^{\pw}=\sum_{\eta=1}^{n}X_{\eta}x_{\eta}^{\pw}-E^{\pw}$, from equation \ref{eq:partition_function} and equation \ref{eq:gen-boltzmann}, we have
\begin{equation}
Z=\sum_\omega\exp\left(\beta\gamma^{\pw}\right)
\end{equation}
and
\begin{equation}
\Pr(\omega)=\frac{\exp\left(\beta\gamma^{\pw}\right)}{Z}\label{eq:thm2pr}
\end{equation}
Therefore
\begin{equation}
\frac{\beta}{Z}\frac{\partial Z}{\partial \beta} = \frac{\beta}{Z}\sum_\omega \gamma^{\pw}\exp\left(\beta\gamma^{\pw}\right) = \sum_\omega\beta \gamma^{\pw}\Pr(\omega).\label{eq:1overzpzpbeta}
\end{equation}
From equation \ref{eq:thm2pr}, we have
\begin{equation}
\beta\gamma^{\pw} = \log Z + \log\Pr\pw.\label{eq:gamma-omega}
\end{equation}
Substitute equation \ref{eq:gamma-omega} and equation \ref{eq:1overzpzpbeta} back to equation \ref{eq:pjpbeta}, we have
\begin{equation}
\frac{\partial J}{\partial \beta} = \frac{1}{\beta^2}\sum_\omega\Pr\pw\log\Pr\pw.\label{eq:pjpbetaeqsumomega}
\end{equation}
Combine equation \ref{eq:seqpjpbata} and equation \ref{eq:pjpbetaeqsumomega}, we have
\begin{equation}
S = -k_B \sum_\omega\Pr\pw\log\Pr\pw.
\end{equation}
This completes the proof.
\end{proof*}

\section{\label{sec:lemmas}Lemmas and their proof}
\begin{lem}
\label{f(abcd)}For a function of 4 variables $f\left(a,b,c,d\right)$,
if
\begin{equation}
a\left.\frac{\partial f}{\partial b}\right|_{acd}+c\left.\frac{\partial f}{\partial d}\right|_{abc}=0,
\end{equation}
then there exists a function $g$ such that $f\left(a,b,c,d\right)=g\left(ad-bc,a,c\right)$.
\end{lem}
\begin{proof*}
Let us call $\left(a,b,c,d\right)$ the old coordinates and define
new coordinates $\left(u,v,w,x\right)$ such that
\begin{equation}
\left\{ \begin{array}{c}
u=a\\
v=c\\
w=ad-bc\\
x=ad+bc
\end{array}\right.
\end{equation}
then the reverse transformation is
\begin{equation}
\left\{ \begin{array}{c}
a=u\\
c=v\\
d=\frac{w+x}{2u}\\
b=\frac{x-w}{2v}
\end{array}\right.
\end{equation}
Evaluating partial derivatives in new coordinate, we have
\begin{eqnarray}
\left.\frac{\partial f}{\partial x}\right|_{uvw} & = & \left.\frac{\partial f}{\partial d}\right|_{abc}\cdot\frac{1}{2a}+\left.\frac{\partial f}{\partial b}\right|_{acd}\cdot\frac{1}{2c}\nonumber\\
& = & \frac{1}{2ac}\left(c\left.\frac{\partial f}{\partial d}\right|_{abc}+a\left.\frac{\partial f}{\partial b}\right|_{acd}\right)=0
\end{eqnarray}
that is, $f$ does not depend on $x$. Therefore, it is a function
of only $u,v,$ and $w$.
\end{proof*}
\begin{lem}
\label{fgc-constant}Let $f$, $g$, and $C$ be functions, and $x,y,a,$ and $b$
be variables. Then $f\left(ax+by,x,y\right)=C\left(a,b\right)\cdot g\left(ax+by,x,y\right)$
implies that $C\left(a,b\right)$ is a constant that does not depend
on $a$ and $b$.
\end{lem}
\begin{proof*}
For fixed $x$ and $y$, the set of all possible values of $\left(a,b\right)$
that has $ax+by=z$ is a line, where $z$ denotes a constant. When $x,y,$
and $ax+by$ are all fixed, the values of $f$ or $g$ does not change. Therefore, $C\left(a,b\right)$ must also be a constant on that line. This
is true for all values of $x,y,$ and $z$, that is, $C\left(a,b\right)$ is a constant on all possible lines. Since different lines cross, then
$C\left(a,b\right)$ must be a constant that does not depend
on $a$ and $b$.
\end{proof*}
\begin{lem}
\label{temp-scale}If we scale the temperature and entropy by $\frac{1}{\alpha}$ and $\alpha$, respectively, we do not change any physics.
\end{lem}
\begin{proof*}
Let us begin our proof by reviewing how the BG theory of equilibrium SM is built in textbooks. We start the procedure by defining the entropy
of the microcanonical ensemble as $S=k_{B}\log\left|\Omega\right|$ , and we establish
a system in thermal equilibrium with a reservoir that defines $T$.
Thus, the number of the microstates of the reservoir $\left|\Omega_{r}\right|$ is given by 
\begin{equation}
\left|\Omega_{r}\right|=\exp\left(S_{r}/k_{B}\right)\label{eq:lemma-omegar}.
\end{equation}
By taking the power series of $S_{r}$ at $E_{\text{total}}$ with respect
to the energy of the system $E_{s}$ to the first order,
we get that
\begin{equation}
S_{r}\left(E_{\text{total}}-E_{S}\right)=S_{r}\left(E_{\text{total}}\right)-\frac{\partial S_{r}}{\partial E_{r}}\cdot E_{S}\label{eq:lemma-sr}
\end{equation}
The application of the first law of thermodynamics gives that 
\begin{equation}
\frac{\partial S_{r}}{\partial E_{r}}=\frac{1}{T}\label{eq:lemma-dsde}
\end{equation}
By combining equations \ref{eq:lemma-omegar}, \ref{eq:lemma-sr}, and \ref{eq:lemma-dsde},
we obtain the following Boltzmann distribution:
\begin{equation}
\left|\Omega_{r}\right|=C\cdot\exp\left(-\frac{E_{s}}{k_{B}T}\right)\label{eq:lemma-oldboltzmann}
\end{equation}
where $C=\exp\left(S_{r}\left(E_{\text{total}}\right)\right)$ denotes
a constant that does not depend on $E_{S}$. In the above procedure,
the temperature scale is introduced by defining entropy as $S=k_{B}\log\left|\Omega\right|$.
The constant $k_{B}$ gets propagated along the logic
chain and determines the temperature scale together with the first
law of thermodynamics.

If we instead had started by defining $S'=\alpha k_{B}\log\left|\Omega\right|$, then the following equation can be obtained by applying the same logic:
\begin{equation}
\left|\Omega_{r}\right|=\exp\left(\frac{S_{r}'}{k_{B}\alpha}\right).
\end{equation}
In this case, the first law of thermodynamics yields
\begin{equation}
\frac{\partial S_{r}'}{\partial E_{r}}=\frac{1}{T'}.
\end{equation}
where $T'$ denotes the temperature in the new scale. Thus, the Boltzmann distribution
in the new scale will look like
\begin{equation}
\left|\Omega_{r}\right|=C\cdot\exp\left(-\frac{E_{s}}{k_{B}\alpha T'}\right)\label{eq:lemma-newboltzmann}.
\end{equation}
Temperature scales are artificial, while probabilities are physical.
Therefore, equation \ref{eq:lemma-newboltzmann} must match with equation \ref{eq:lemma-oldboltzmann}.
To prove this, we utilize the first law of thermodynamics to obtain the relationship between $T$ and $T'$:
\begin{equation}
\left\{ \begin{array}{c}
dU=TdS-pdV\\
dU=T'dS'-pdV\\
S'=\alpha S
\end{array}\right.\Rightarrow T'=\frac{T}{\alpha}\label{eq:TT'}.
\end{equation}
By substituting equation \ref{eq:TT'} into equation \ref{eq:lemma-newboltzmann}, we obtain an exact match with equation \ref{eq:lemma-oldboltzmann}. This completes this
proof.

\end{proof*}

\section{Discussion}

In this article, we discussed what the basic components of an ensemble theory are. These basic components include the set of microstates, the different roles that different thermodynamic state functions play (these roles include:  parameters determining $\Omega$, parameters determining the probability density, quantities associated with random variables, and other statistical quantities), the probability density, and the set of rules connecting each random variable with its corresponding thermodynamic state functions. We showed how the mathematical form of these basic components, together with some additional assumptions, can be used to derive the generalized Boltzmann distribution. This derivation only uses the mathematical form of these components and the consistency with thermodynamics; no prior distribution is required.

Since definition \ref{def:system} is based on the textbook approach of equilibrium thermodynamics, the primary purpose of this article is to reveal the internal structure of the classical theory of equilibrium thermodynamics and statistical mechanics. However, the author would like to remind the reader that definition \ref{def:system}, the three assumptions, and the proof of theorem \ref{fake-dist-thm} and theorem \ref{thm:entropy} are all about mathematical forms and contains little about the physical interpretation of these mathematics. As a result, the argument in this article can be naturally extended to non-equilibrium theories that share the same mathematical form.

As a side note, the reader can easily verify that, if assumption \ref{ass:avg} is replaced with the corresponding equation in \cite{Curado1991}, the conclusion of theorem \ref{fake-dist-thm} (i.e. equation \ref{eq:gen-boltzmann}) is replaced with the $q$-distribution, the equation \ref{eq:G'=g} is replaced with $G'=g^{q}$, and the conclusion of theorem \ref{thm:entropy} (i.e. equation \ref{eq:entropy}) is replaced with the $q$-entropy, our arguments still hold. This means that our approach can also be used to obtain the Tsallis statistics.

\section{Acknowledgment}

The author thanks Constantino Tsallis for the discussion on the connection with the Tsallis statistics. The authors would like to thank Enago (www.enago.com) for the English language review.



\bibliography{main}

\end{document}